\documentclass[sigconf]{acmart}
\usepackage{graphicx}
\usepackage{hyperref}
\usepackage{xcolor}
\usepackage{soul}
\usepackage{lipsum}
\usepackage{subcaption}
\usepackage{array}

\usepackage{enumitem}

\copyrightyear{2024}
\acmYear{2024}
\setcopyright{rightsretained}
\acmConference[SIGSPATIAL '24]{The 32nd ACM International Conference on Advances in Geographic Information Systems}{October 29-November 1, 2024}{Atlanta, GA, USA}
\acmBooktitle{The 32nd ACM International Conference on Advances in Geographic Information Systems (SIGSPATIAL '24), October 29-November 1, 2024, Atlanta, GA, USA}
\acmDOI{10.1145/3678717.3691319}
\acmISBN{979-8-4007-1107-7/24/10}
\newcommand{\GitHubProposedApproach}[1]{\url{https://github.com/onspatial/generate-mobility-dataset#1}}

\newcommand{\reffig}[1]{Figure \ref{#1}}

\newcommand{\reftab}[1]{Table \ref{#1}}

\renewcommand\footnotetextcopyrightpermission[1]{}  

\usepackage{color}

\begin{document}

\title{The Patterns of Life Human Mobility Simulation}
\author{Hossein Amiri}
\orcid{0000-0003-0926-7679}
\affiliation{%
    \institution{Emory University, USA}
    \city{}
    \country{}
}
\email{hossein.amiri@emory.edu}

\author{Will Kohn}
\orcid{0009-0003-7265-9082}
\affiliation{
    \institution{Emory University, USA}
    \city{}
    \country{}
}
\email{will.kohn@emory.edu}

\author{Shiyang Ruan}
\orcid{0000-0002-0279-4719}
\affiliation{%
  \institution{George Mason University, USA}
  \city{}
   \country{}
}
\email{sruan@gmu.edu}

\author{Joon-Seok Kim}
\orcid{0000-0001-9963-6698}
\affiliation{%
  \institution{Emory University, USA}
    \city{}
    \country{}
}
\email{joonseok.kim@emory.edu}

\author{Hamdi Kavak}
\orcid{0000-0003-4307-2381}
\affiliation{%
  \institution{George Mason University, USA}
    \city{}
    \country{}
}
\email{hkavak@gmu.edu}

\author{Andrew Crooks}
\orcid{0000-0002-5034-6654}
\affiliation{%
  \institution{University at Buffalo, USA}
    \city{}
    \country{}
}
\email{atcrooks@buffalo.edu}

\author{Dieter Pfoser}
\orcid{0000-0001-9197-0069}
\affiliation{%
  \institution{George Mason University, USA}
    \city{}
    \country{}
}
\email{dpfoser@gmu.edu}

\author{Carola Wenk}
\orcid{0000-0001-9275-5336}
\affiliation{%
  \institution{Tulane University, USA}
    \city{}
    \country{}
}
\email{cwenk@tulane.edu}

\author{Andreas Z{\"u}fle}
\orcid{0000-0001-7001-4123}
\affiliation{%
    \institution{Emory University, USA}
    \city{}
    \country{}
}
\email{azufle@emory.edu}

\renewcommand{\shortauthors}{Amiri, Ruan, Kim et al.}

\renewcommand{\shortauthors}{Amiri, et al.}
\begin{abstract}
    We demonstrate the Patterns of Life Simulation to create realistic simulations of human mobility in a city. This simulation has recently been used to generate massive amounts of trajectory and check-in data. Our demonstration focuses on using the simulation twofold: (1) using the graphical user interface (GUI), and (2) running the simulation headless by disabling the GUI for faster data generation. We further demonstrate how the Patterns of Life simulation can be used to simulate any region on Earth by using publicly available data from OpenStreetMap. Finally, we also demonstrate recent improvements to the scalability of the simulation allows simulating up to 100,000 individual agents for years of simulation time.
During our demonstration, as well as offline using our guides on GitHub, participants will learn: (1) The theories of human behavior driving the Patters of Life simulation, (2) how to simulate to generate massive amounts of synthetic yet realistic trajectory data, (3) running the simulation for a region of interest chosen by participants using OSM data, (4) learn the scalability of the simulation and understand the properties of generated data, and (5) manage thousands of parallel simulation instances running concurrently.

\end{abstract}

\ccsdesc[500]{Information systems~Geographic information systems }
\ccsdesc[500]{Information systems~Location based services }

\keywords{ Patterns of Life, Simulation, Trajectory, Dataset, Customization}

\maketitle

\section{Introduction}
\label{sec:introduction}

\begin{figure*}[!ht]
    \centering
    \includegraphics[trim = 0cm 0.5cm 2cm 1cm, clip, width=1\linewidth,height=8.5cm]{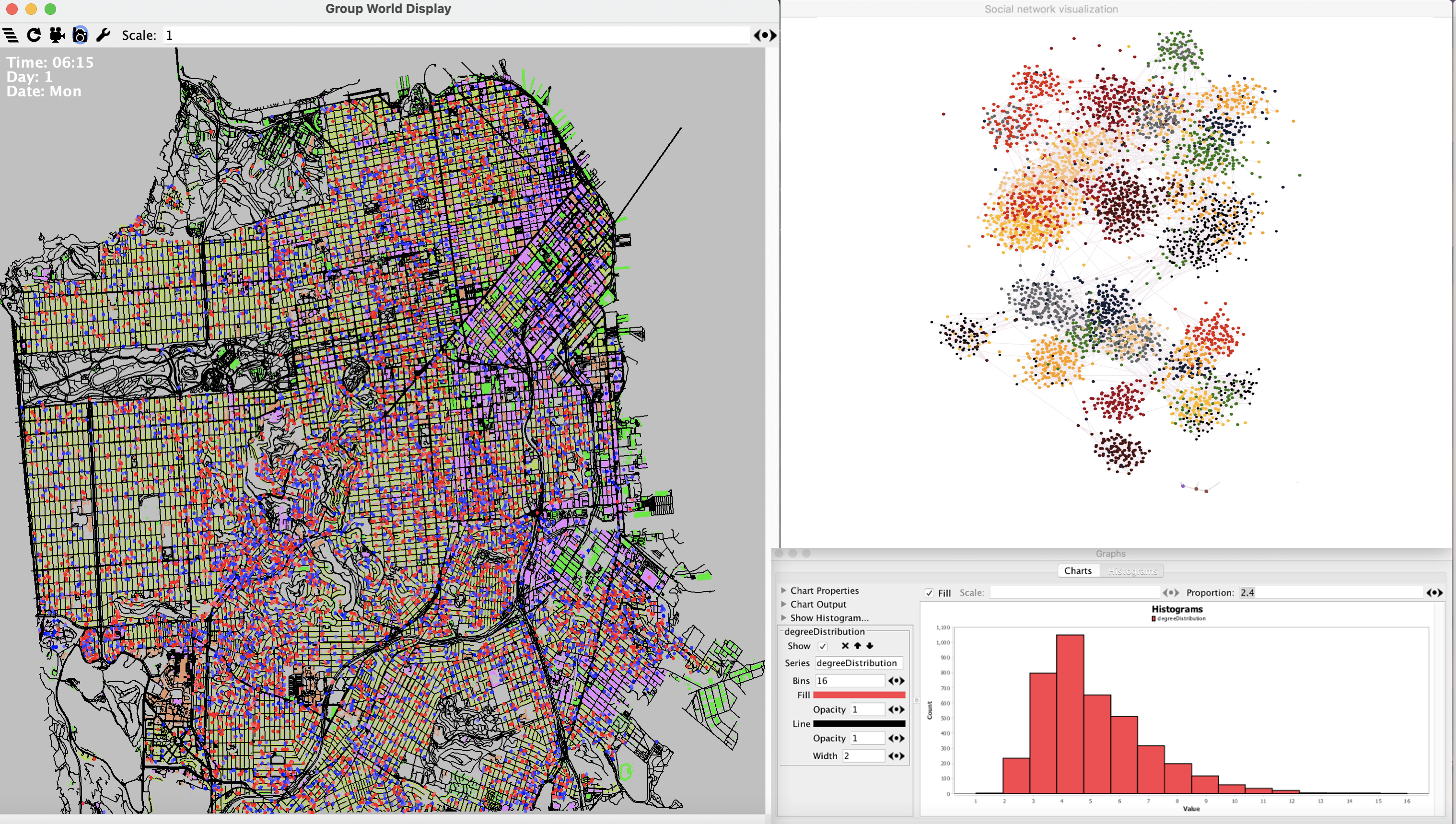}
    \caption{A screenshot of the graphical user interface of the Patterns of Life Simulation. The GUI shows the map and the movements of agents on the left side and the social network of agents and their statistical properties on the right side. }
    \label{fig:pol-gui}
\end{figure*}

Trajectory data~\cite{amiri2023massive} captures frequently measured location sequences of objects such as humans~\cite{zheng2010geolife}, vehicles~\cite{yuan2011t}, and animals\cite{calenge2009concept}. Trajectory data enables a deeper analysis of human behaviors~\cite{toch2019analyzing,zhu2024generic}, their mobility~\cite{mokbel2018mobility,mokbel2022mobility}, enhancing infectious disease modeling~\cite{kohn2023epipol}, and facilitating urban mobility studies~\cite{zhu2024synmob,barbosa2018human,gonzalez2008understanding}. Additionally, it is instrumental in traffic analysis~\cite{chen2022lane}, the detection of anomalies~\cite{zhang2023large,zhang2024transferable,liu2024neural,amiri2024urban} and urban planning~\cite{isaacman2012human,zhao2016urban}. Despite the plethora of applications for analyzing trajectory data, research is hindered by a lack of available trajectory data sets. The largest and most commonly used trajectory dataset of individual human trajectories is GeoLife~\cite{zheng2010geolife} which captures only 182 individual human users where  most of these users are captured only for a few hours or even minutes~\cite{kim2020location}.

To fill this gap and enable research on large trajectory datasets, our team has developed the Patterns of Life Simulation~\cite{zufle2023urban,kim2019simulating} which was created in Java leveraging the MASON agent-based modeling toolkit \cite{LukeMASONSimulationToolkit2018}. The simulation creates realistic worlds to support in-silico~\cite{zufle2024silico} experiments to study hypothesis without involving real populations. The simulation uses Maslowian~\cite{maslow1943theory} needs of agents (i.e., people) which drive agents' actions and behaviors like going home to find shelter; working to make money; going to restaurants to eat; and visiting recreational sites to meet friends. Many more facets of the logic driving the behavior of agents in the Patterns of Life simulation can be found in~\cite{zufle2023urban}. \reffig{fig:pol-gui} shows a screenshot of the graphical user interface (GUI) used to show the movement of agents across a study region and their social network.
%
We used the Patterns of Life simulation to generate large sets of check-in data~\cite{kim2020location} and large sets of trajectory data~\cite{amiri2023massive}. However, these datasets pertain only to a small set of pre-selected study regions. This demonstration  will show how the Patterns of Life Simulation can be adapted by users for their own purpose and their own study region. Specifically, this demonstration explains how to:
\begin{itemize}
    \item Run the Patterns of Life Simulation with the GUI to visually explore the mobility patterns of a region
    \item Run the Patterns of Life Simulation headless (without GUI) for large-scale data generation
    \item Adapt the simulation to any region in the world using OpenStreetMap~\cite{overpass_turbo} data
    \item Showcase  recent scalability improvements  allowing to simulate hundreds of thousands of agents
    \item Demonstrate a framework that supports concurrently running thousands of instances of the simulation.
\end{itemize}


\section{Demonstration}
\label{sec:methodology}
The demonstration will be organized as follows:
Section \ref{sec:POL_GUI} provides an overview of the Patterns of Life Simulation using GUI, and Section~\ref{sec:POL_Headless} demonstrates how to use the simulation for data generation with the GUI. Then Section~\ref{sec:Maps} describes how OpenStreetMap data can be used to change the simulation to any region of interest in the World. Section~\ref{sec:scalability} details the changes we have made to increase the scalability of the simulation, and Section~\ref{sec:parallel} demonstrates a framework to run many simulations in parallel for very large-scale trajectory data generation. 
In this demonstration, we utilized what we call the ``Vanilla'' simulation as a baseline and is located in the following GitHub repository: \url{{https://github.com/gmuggs/pol}}. All the improvements, optimization, and documentation to reproduce this demonstration are found at~\GitHubProposedApproach{}


\subsection{Patterns of Life (POL) Demonstration: GUI}\label{sec:POL_GUI}\vspace{-0.0cm}
For the demonstration to be presented at SIGSPATIAL~'24, we will run the Patterns of Life Simulation live to generate trajectory data. We have prepared several maps, including those of Atlanta, USA; Beijing, China; Berlin, Germany; and San Francisco, USA. By default, we will use the Atlanta map using 1000 agents for this demonstration. We choose such as small number of agents to avoid waiting times during the demonstration for simulation initialization (we will come back to run time of the simulation in Section~\ref{sec:scalability}). First, we will demonstrate a single run of the simulation using the GUI to explain how the simulation works, how agents move across the map, and how they interact with each other to form social networks as seen in Figure~\ref{fig:pol-gui}. While the simulation is running, we will give a brief overview of the theories of social behavior guiding agents' behavior including Maslow's Hierarchy of Needs~\cite{maslow1943theory} and the Theory of Planned Behavior~\cite{ajzen1991theory}. For the live-audience at SIGSPATIAL~'24 we will include details of these theories on a poster that will be exhibited next to our demonstrator. We will also briefly explain how the social network (on the right of the GUI shown in Figure~\ref{fig:pol-gui}) is formed and evolves. The interested reader of this paper may find these details in our simulation description~\cite{zufle2023urban}. After simulating for about one minute (wall-clock time), about ten days of simulation time will have been generated. We will show the generated log files including trajectories, check-ins, and time-series of agent states.

\subsection{POL Demonstration: Headless}\label{sec:POL_Headless}
\begin{figure}
    \centering
    \includegraphics[ width=1.0\linewidth]{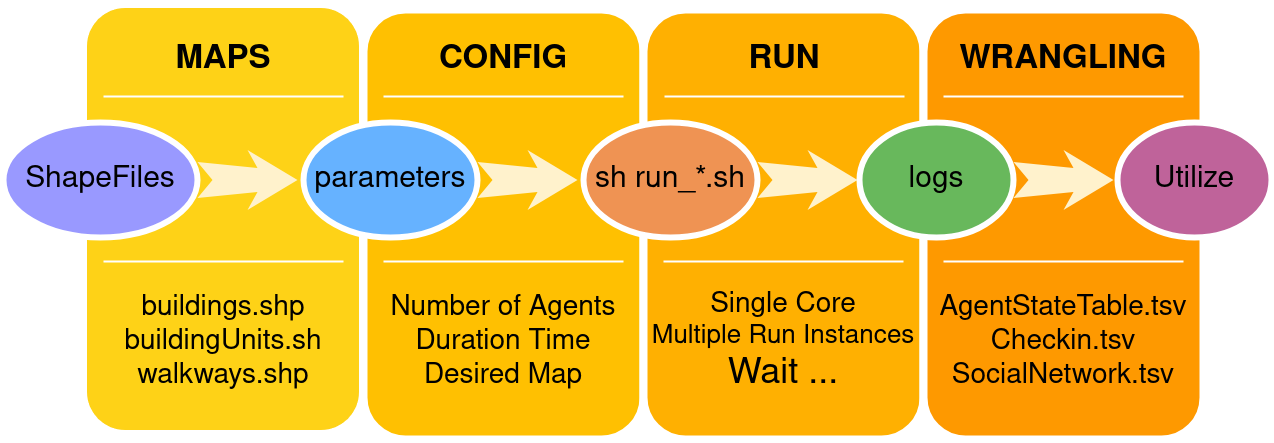}
    \caption{Steps to generate the one trajectory dataset.}
    \label{fig:run-simulation}
    \vspace{-0.4052cm}
\end{figure}

To use the Patterns of Life Simulation for large-scale data generation without having the GUI as a computational bottleneck, we also show how the simulation can be run headless. The overall process to run the simulation and generate a single trajectory dataset is outlined in~\reffig{fig:run-simulation}.
After generating the maps and configuring the simulation parameters, the simulation is started using a bash script using the pre-compiled \texttt{jar} file of the Patterns of Life simulation. The simulation generates the logs in the specified directory. The logs contain information about the agents' movements, interactions, social network, check-in, and other data that can be specified for specific applications or disabled for efficiency. 

\begin{figure}
    \centering
    \includegraphics[width=0.99\linewidth]{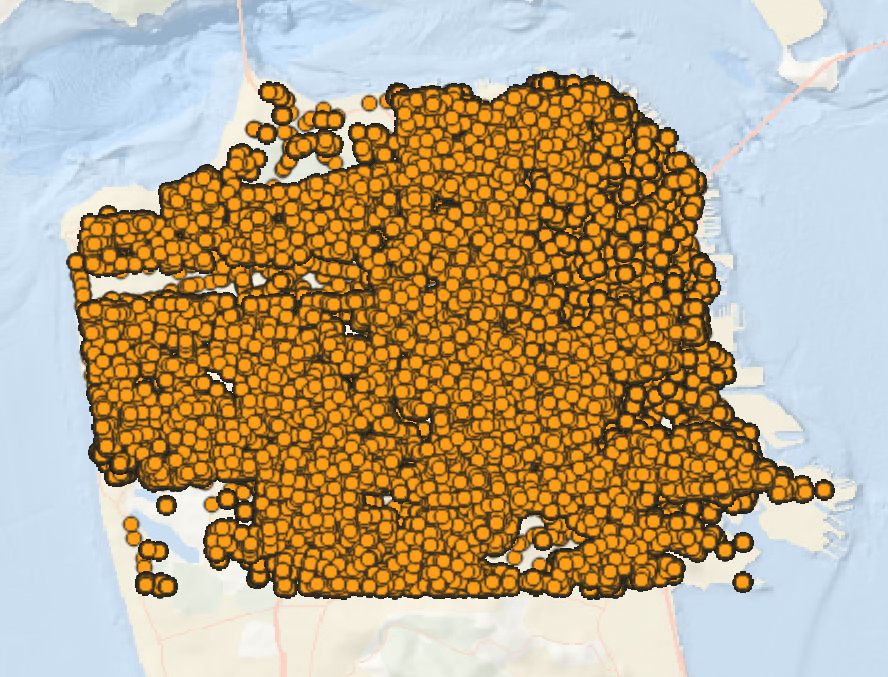}
    \caption{Example visualization of the generated dataset for San Francisco, California, USA}
    \label{fig:example_data}
    \vspace{-0.4052cm}
\end{figure}

Using a Bash script on Linux, we will demonstrate a use case that involves running a single simulation instance with predefined parameters to showcase the data generation process for that instance. Furthermore, we will explore the creation and customization of log files during this stage, using both the simulation source code and the provided data processing scripts. To demonstrate the data visually, our poster will also show a simulated data set such as Figure~\ref{fig:example_data} which shows a downsampled version of generated trajectories data in San Francisco. 
We will illustrate how to alter the simulation's source code to integrate custom features and subsequently compile this modified code into a JAR file.

\subsection{Adapting POL to Simulate other Region}\label{sec:Maps}
\begin{figure}
    \centering
    \includegraphics[trim = 2.5cm 0 0 2.5cm, clip, width=0.95\linewidth]{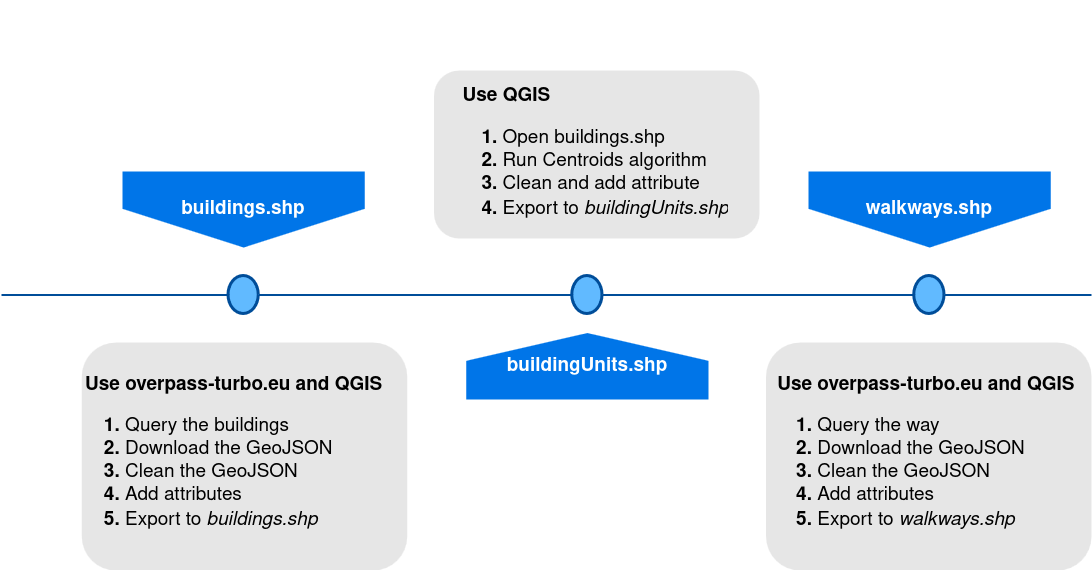}
    \vspace{-0.251cm}
    \caption{Steps to generate a new map for the simulation.}
    \label{fig:new-map}
    \vspace{-0.20651cm}
\end{figure}

Then, we will ask people in the audience which region of interest, anywhere in the world, they would like to simulate. We will obtain map data for this region from OpenStreetMap (OSM), transform the data into the format required by the simulation, run the simulation on this new region, and explain how the simulation uses information about building footprints as well as information on residential and commercial areas to initialize the simulation. 
Specifically, to generate a new map for the simulation, we follow the steps outlined in~\reffig{fig:new-map} to create the spatial layers required for the simulation environment. Specifically, the simulation relies on three distinct shapefiles: \texttt{buildings.shp}, \texttt{buildingUnits.shp}, and \texttt{walkways.shp}. The \texttt{buildings.shp} contains the footprints of buildings and information on their usage as residential or commercial. While OSM does not include building usage information for every building, it has been shown that building usage can be accurately predicted from other building features~\cite{atwal2022predicting}. This file is created using the Overpass API for OSM and then processed in QGIS to filter and assign attributes that define building type. The \texttt{buildingUnits.shp} file is then generated from the buildings.shp, by creating smaller individual units (apartments, shops) within a building, thereby facilitating more granular simulations of movement and interaction within the building premises.
In the same way, the \texttt{walkways.shp} file is created to map out the 
transportation network for agents to move between buildings. Once these files are generated and placed in the corresponding folders, the Patterns of Life Simulation is started with the new study region. For readers unable to attend the demonstration in-person, we document instructions how OSM data can be quickly obtained and used in the simulation in our GitHub repository at \GitHubProposedApproach{/blob/main/documentation/maps.md}


\subsection{Large-Scale Data Generation}\label{sec:scalability}

Initial profiling of the original Patters of Life Simulation, facilitated by VisualVM~\cite{javaprofiler}, enabled the identification of critical performance bottlenecks. A key finding was the importance of initial parameter settings, such as the agent's starting financial balance. During early simulation stages, agents predominantly struggled with paying their rent. This caused agents to frequently change jobs and home locations in an attempt to satisfy their Financial Need (i.e., earn money). This behavior created a substantial computational bottleneck. In the updated version used for this demonstration, we limited the agent's ability to re-evaluate their home and job to only once-per-day, rather than every five-minute tick. In addition, we changed agents' decision-making to use Euclidean distance to find the nearest places of interest rather than network distance. While this is a change to the agents' logic, we argue that this change is realistic, as humans do not run a Dijkstra single-source shortest-path search to find nearby places to visit, but rather use an approximation based on the human's mental map of the area~\cite{manley2015shortest}. 
These modifications, coupled with general code cleanup and optimization of the logging system, markedly improved simulation performance. We also addressed compile-time warnings and disabled journal record updates at each simulation step to streamline execution. Additionally, we enhanced output readability, for instance, by converting the time display from milliseconds to a more comprehensible format. These adjustments not only alleviated the identified bottlenecks but also improved the overall functionality and reliability of the simulation.

We conducted a comparative analysis of the initialization and simulation time  between the original (``Vanilla'') and the improved (``Improved'') version of the Patters of Life Simulation. The results are presented in \reftab{tab:comparison_time}. The new version is significantly faster than the Vanilla version in both initialization and simulation time. For instance, when simulating 1000 agents, the ``improved'' version is 3.8 times faster in initialization and 8.1 times faster in simulation time compared to the ``vanilla`` version. The performance improvement is more pronounced as the number of agents increases. For example, when simulating 150,000 agents, it was not reasonable to run the Vanilla version due to the long initialization and simulation time. In contrast, the "Improved" version can handle this number of agents with a reasonable time frame. An example of generated data can be found in~\cite{amiri2023massive} we used the framework to generate a dataset with various number of agents and maps for different regions. Since we won't have time to regenerate these very large datasets at SIGSPATIAL, we will instead explore already-generated datasets which include billions of check-ins, trillions of social links, and tens of trillions of trajectory points. We will note that these numbers are many orders of magnitude larger than any existing datasets. The reader can find more details in our Data \& Resource Paper~\cite{amiri2023massive}.

\subsection{Parallelization of POL}\label{sec:parallel}
Since the simulation uses one single core, other simulation instances can be run in parallel to generate more data.
We developed tools to run simulations in parallel, parameterize the simulations for specific use cases, and process logs to format output data in a customized format. The parallelization framework can operate using either fork-join or job-queue approaches. In the fork-join model, all processes must wait for the slowest one, whereas the job-queue model does not wait and ensures maximum parallelism is achieved. The fork-join approach is suitable when the next generation in a parallel world requires information from its predecessors, such as in genetic algorithms. Conversely, the job-queue approach is beneficial for studying varying parameter data without dependencies. Upon completion of the simulations, data processing techniques can be employed to concatenate and pre-process the logs for more in-depth studies, such as plotting them or run machine learning algorithms on the output. The source code for the enhanced simulation, parallelization framework, and data-processing tools is available online at~\GitHubProposedApproach{}.
We will demonstrate how to set up a large number of simulations, allow the algorithms to run simulations in parallel, and generate data. Additionally, we will discuss how to calibrate the parameters to ensure that simulation data realistically models the real-world.

\begin{table}
    \centering
    \begin{tabular}{|l|c|c|c|c|}
        \hline
        Time    & \multicolumn{2}{c|}{Initialization} & \multicolumn{2}{c|}{Simulation}                           \\
        \hline
        Agents  & Vanilla                             & Improved                           & Vanilla    & Improved      \\
        \hline
        1000    & 33.22 s                             & 8.66 s                          & 10.3 min   & 1.27 min   \\
        \hline
        5000    & 4.61 min                            & 48.83 s                         & 3.69 hour  & 5.68 min   \\
        \hline
        10,000  & 13.87 min                           & 3.27 min                        & 15.8 hour  & 14.73 min  \\
        \hline
        15,000  & 24.12 min                           & 8.74 min                        & 35.45 hour & 32.56 min  \\
        \hline
        100,000 & DNF                                  & 4.37 hour                       & DNF         & 9.0 hour   \\
        \hline
        150,000 & DNF                                  & 9.65 hour                       & DNF         & 19.29 hour \\
        \hline
    \end{tabular}
    \caption {Comparative Analysis of Initialization and Simulation Time  over a Ten-Day Simulation Period}
    \label{tab:comparison_time}
    \vspace{-1.1cm}
\end{table}






\section{Conclusion}
\label{sec:conclusion}
Human mobility data is a crucial component in various research fields, including urban planning, epidemiology, and transportation. The availability of large-scale human mobility data has enabled researchers to develop models and tools to analyze and predict human behavior. However, generating such data is challenging due to the complexity of human mobility patterns and the need for large-scale data to develop accurate models. In this paper, we present a framework for generating large-scale human mobility data using an existing agent-based simulation. We demonstrate how to optimize the simulation code, generate new maps, run the simulation, and process the logs to generate the final data. We also show how to run multiple simulations in parallel to generate large-scale data efficiently. We present the results of our framework, including the performance improvement of the optimized simulation code and the generated data. The framework is flexible and can be adapted to different research needs by customizing the simulation parameters and data processing steps. The source code for the enhanced simulation, parallelization framework, and data-processing tools is available online at~\GitHubProposedApproach{}.\vspace{-0.15cm}

\bibliographystyle{ACM-Reference-Format}
\bibliography{main}

\end{document}